\documentclass[useAMS,usenatbib,aas_macros]{mn2e}

\usepackage{lscape}
\usepackage{graphicx}
\usepackage{colortbl}
\usepackage{amssymb}
\usepackage{natbib}
\usepackage{times}
\usepackage{aas_macros}

\newcommand{\msun}{\rm M_\odot}

\newcommand{\bhbulge}{M$_{\rm BH}-$M$_{\rm bulge}\ $}
\newcommand{\Mbh}{M_{\rm{BH}}}

\usepackage{color}

\def\simpropto{\lower.2ex\hbox{$\; \buildrel \propto \over \sim \;$}}
\def\ltsim{\lower.5ex\hbox{$\; \buildrel < \over \sim \;$}}
\def\gtsim{\lower.5ex\hbox{$\; \buildrel > \over \sim \;$}}

\begin{document}

\title[Unravelling obese black holes in the first galaxies]{Unravelling obese black holes in the first galaxies}

\author[B. Agarwal, et al.]{Bhaskar Agarwal$^1$\thanks{E-mail:
agarwalb@mpe.mpg.de}, Andrew J. Davis$^1$ Sadegh Khochfar$^{1,3}$, Priyamvada Natarajan$^2$,
\newauthor James S. Dunlop$^3$\\
$^1$Max-Planck-Institut f{\"u}r extraterrestrische Physik,
Giessenbachstra\ss{}e, 85748 Garching, Germany\\
$^2$Department of Astronomy, 260 Whitney Avenue, Yale University, New Haven, CT, USA\\
$^3$Institute for Astronomy, University of Edinburgh, Royal Observatory, Edinburgh, EH9 3HJ}


\date{00 October 2012}
\pagerange{\pageref{firstpage}--\pageref{lastpage}} \pubyear{0000}
\maketitle

\label{firstpage}


\begin{abstract}
We predict the existence and observational signatures of a new class of objects 
that assembled early, during the first billion years of cosmic time: Obese Black-hole Galaxies (OBGs). 
OBGs are objects in which the mass of the central black hole initially exceeds 
that of the stellar component of the host galaxy, and the luminosity from black-hole accretion 
dominates the starlight. Conventional wisdom dictates that the first galaxies 
light up with the formation of the first stars; we show here that, in fact, there 
could exist a population of astrophysical objects in which this is not the case. 
From a cosmological simulation, we demonstrate that there are sites where star formation is 
initially inhibited and direct-collapse black holes (DCBHs) form
due to the photo-dissociating effect of Lyman-Werner radiation 
on molecular hydrogen. We show that the formation of OBGs is inevitable,
because the probability of finding the required extra-galactic environment and the 
right physical conditions in a halo conducive to DCBH formation is quite high in the early universe. 
We estimate an OBG number density of 0.009 Mpc$^{-3}$ at $z\sim8$ and 0.03 Mpc$^{-3}$ at $z\sim6$.  Extrapolating 
from our simulation volume, we infer that the most luminous quasars detected 
at $z\geq6$ likely transited through an earlier OBG phase. Following the growth 
history of DCBHs and their host galaxies in an evolving dark matter halo shows that 
these primordial galaxies start off with an over-massive BH and acquire their 
stellar component from subsequent merging as well as in-situ star formation. 
In doing so, they inevitably go through an OBG phase dominated by the accretion 
luminosity at the Eddington rate or below, released from the growing BH. 
The OBG phase is characterised by an ultra-violet (UV) spectrum $f_{\lambda} \propto \lambda^{\beta}$ 
with slope of $\beta \sim -2.3$ and the absence of a Balmer Break. 
OBGs should also be spatially unresolved, and are expected to be brighter 
than the majority of known high-redshift galaxies. They could also display broad high-excitation emission lines, as 
expected from Type-I active galactic nuclei (AGN), although the strength of lines such as NV and CIV 
will obviously depend on the chemical enrichment of the host galaxy. OBGs could potentially be revealed via 
{\it Hubble Space Telescope} (HST) follow-up imaging of samples of brighter Lyman-break galaxies 
provided by wide-area ground-based surveys such as UltraVISTA, and should be easily uncovered and studied with instruments aboard the {\it James Webb Space Telescope} (JWST). The discovery and characterization of OBGs would provide important insights into the formation of the first black-holes, and their influence on early galaxy formation.

\end{abstract}


\section{Introduction}
\label{sec:Intro}
It is now well-established that most present-day galaxies harbour a quiescent super-massive 
black hole (SMBH), with a mass approximately one thousandth of the mass of stars in the bulge \citep{Ferrarese:2000p830,Haring:2004p1017}. Such a correlation is strongly suggestive of coupled 
growth of the SMBH and the stellar component, likely via regulation of the gas supply in galactic nuclei from the earliest times \citep[e.g.][see however \citealt{Hirschmann:2010p2277}]{Silk:1998p1456,Haehnelt:2000p1457,
Fabian:2002p1929,King:2003p1930,Thompson:2005p1579,Robertson:2006p2283,
Hopkins:2009p1931,Natarajan:2009p1589,Treister:2011p114}. 

Since the same gas reservoir fuels star formation and feeds the black hole, a 
connection between these two astrophysical processes regulated by the evolving gravitational 
potential of the dark matter halo is arguably expected. However, understanding when and how this interplay commences has been both a theoretical and observational challenge for current theories of structure formation. 
 
In this letter, we explore the formation and evolution of the first 
massive black-hole seeds and the first stars during the earliest epochs 
in order to explore the onset of coupling between the black hole and the stellar component. 
Our calculation incorporates two new physical processes that have only been recently 
recognised as critical to understanding the fate of collapsing gas in the early 
universe. The first is the computation of the Lyman-Werner (LW) radiation 
($11.2 - 13.6 \ \rm eV$) that impacts gas collapse in the first dark-matter haloes 
(as it is able to efficiently dissociate the H$_2$ molecules, thereby 
preventing cooling via molecular hydrogen; e.g. \citet{Haiman:2000p87}). 
The second is the implementation of our growing understanding of the role 
of the angular momentum of the baryonic gas in the collapse process 
\citep{Lodato:2006p375,Davis:2010p139}. 
Including these two processes within 
the context of the standard paradigm of structure formation,  predicts the possible existence of a new class of object in the high-redshift Universe in which black-hole growth commences before, and continues to 
lead the build up of the galaxy stellar population for a significant period of time. We define 
an OBG as a phase in a galaxy's evolution where post DCBH formation, the BH at least initially dominates over the stellar mass and is accreting at a rate sufficiently high enough to outshine the stellar component.

OBGs may provide a natural stage of early black-hole/galaxy evolution en route to the most luminous 
quasars already observed to be in place at $z \simeq 6$ with estimated black-hole masses 
$M_{\rm BH} \simeq 10^9\,{\rm M_{\odot}}$. Observationally, such 
objects should appear similar to moderate-luminosity AGN, but with very low-luminosity 
host galaxies, and low metallicities. 

\begin{figure}
\centering
\includegraphics[width=\columnwidth,trim=-10mm -10mm 5 0]{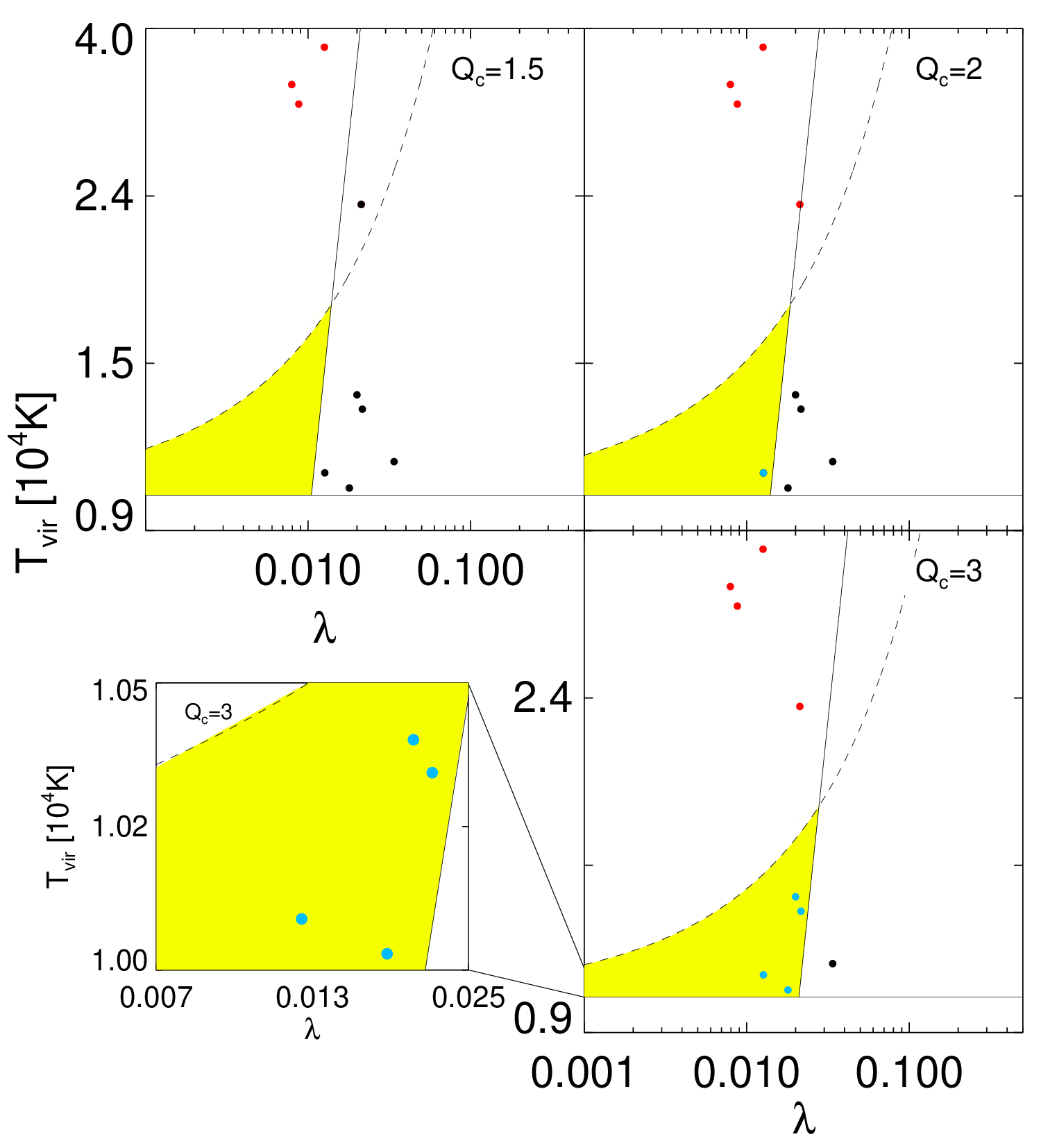}
\caption{The temperature/spin distribution of pristine massive dark-matter 
haloes that are exposed to $J_{\rm LW} \ge J_{\rm crit}$. The virial 
temperature, $T_{\rm vir}$, is plotted against the halo spin parameter, $\lambda$, for all DCBH candidates with Toomre 
stability parameter $Q_c = 1.5, 2, 3$. The nearly-vertical solid curve represents the value of $\lambda_{\rm max}$ 
and all haloes with spin less than this critical value lie to the left of this line (red points), 
whereas those with larger spin lie to the right (black points). The upper limit on the scale-length 
of the hosted disc given the allowed $T_{\rm vir}-\lambda_{\rm max}$ 
combination is marked as the dashed line. Any halo in the yellow region, i.e. below the dashed line and to the left of the 
solid vertical curve will host a DCBH (blue points) in our model. Note that the yellow region shrinks as $Q_c$ decreases, thereby reducing the probability of finding a halo 
that can form a DCBH. Inset: A zoom-in on the $T_{\rm vir}-\lambda$ distribution of the four DCBH candidates in our fiducial case with $Q_{c}=3$.}
\label{Figure1}
\end{figure}


\begin{figure}
\centering
\includegraphics[width= \columnwidth]{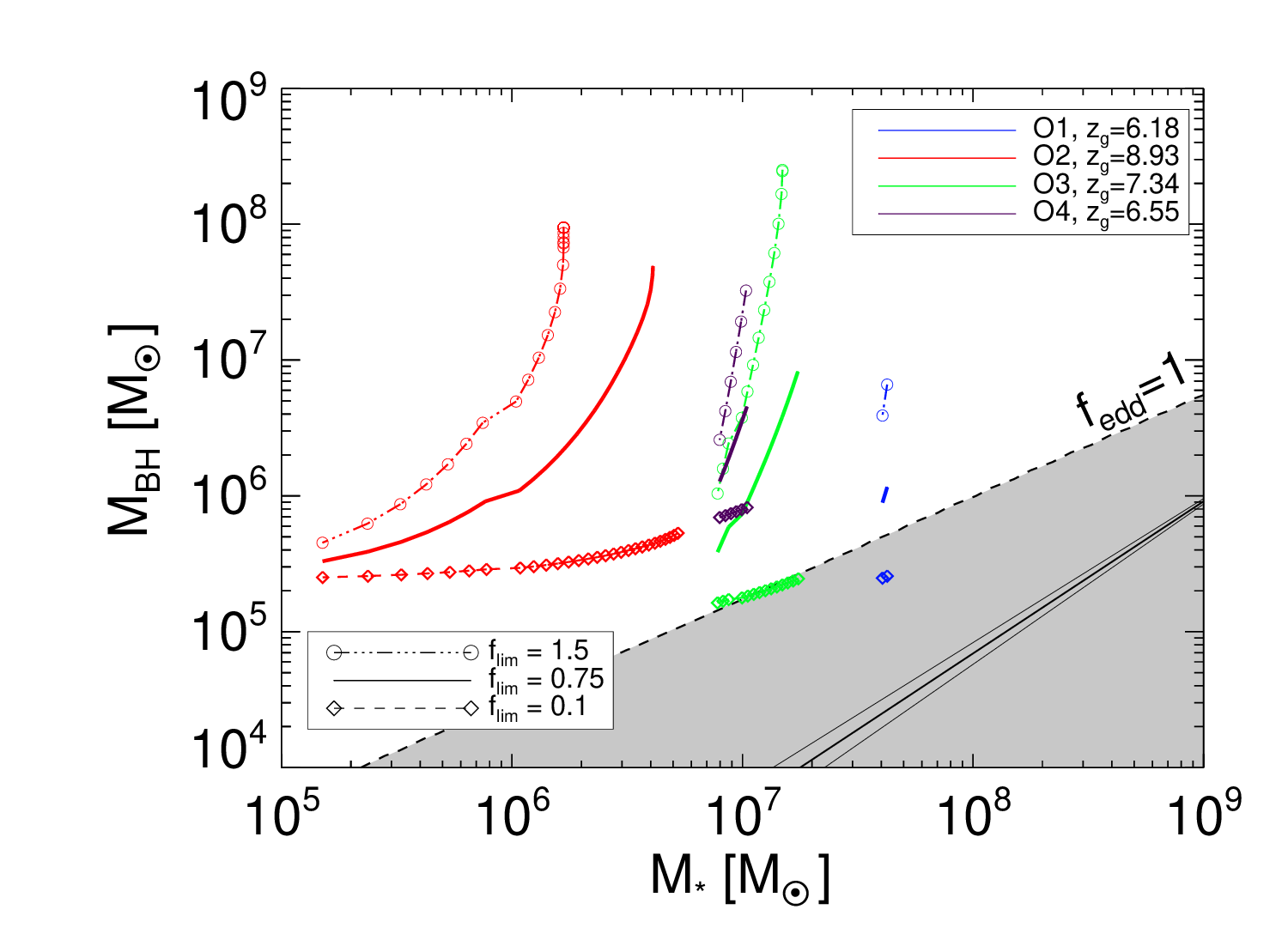}
\caption{Predicted redshift evolution of $\Mbh$ and $M_*$ 
for the four OBGs in our simulation down to $z=6$. We show the tracks for different fractions of the Eddington 
accretion rate $f_{\rm lim}=1.5, 0.75\ \rm (fiducial), 0.1$. The parameter $z_g$ denotes the redshift when the galaxy becomes an OBG and first appears on this plot. 
The shaded portion represents the region where the BH's accretion rate 
would have to be larger than the Eddington limit for the galaxy to qualify as an OBG.  In solid-black lines, 
we show the local \bhbulge relation and the 1--$\sigma$ error in the fit \citep{Haring:2004p1017}. }
\label{Figure2}
\end{figure}

\section{Methodology}
\label{sec:Method}

Our model is a modified version of \cite{Agarwal:2012p2110}, A12 here after, where they identify the sites of 
DCBH formation by calculating the total amount (spatial and global) of LW radiation seen by any given halo within a cosmological 
N-body dark-matter only simulation using a semi-analytic model for the star formation in these haloes. 
The key features of A12 are summarised below:
\begin{enumerate}
\item The DM only N--body simulation is run from $z=30$ to $z=6$ with a box size of $\sim$3.4\,Mpc $h^{-1}$ and DM particle mass of $\rm 6500\ M_{\sun}$$h^{-1}$. This was chosen so that we can resolve a minimum halo mass $\sim 10^5 \ \rm M_{\sun}$ with  20 particles, similar to the minimum halo mass that can host a Pop III star at $z=30$ \citep{Tegmark:1997p937}.
\item Both Pop III and Pop II star formation are allowed, and halo histories are tracked 
in order to determine if a halo is metal free. 
\item Many realisations of the model are run to study the effect of different LW escape fractions, 
Pop II star formation efficiencies, gas outflow rates due to supernova feedback, number of Pop III stars forming per halo 
and reionisation feedback. The results presented here are based on the run with a LW escape 
fraction of 1.0 and a Pop II star-formation efficiency of 0.005 with a burst mode of star formation. We create Pop II stars in a single burst that is placed randomly between the two time steps for which we use the burst mode template (Fig. 7e) from \textsc{starburst99} \citep{Leitherer:1999p112}. Using the burst mode leads to a peak in LW emission at $10^{-42}$ erg/s for a 1 Myr old, $10^6 \msun$ Pop II star cluster which drops to $10^{-38}$ erg/s at 700 Myr.
\item The LW specific intensity in units of $10^{-21}$ erg s$^{-1}$cm$^{-2}$Hz$^{-1}$sr$^{-1}$ , $J_{\rm LW}$ is 
computed self-consistently depending on the type, mass and age of the stellar population, and with two components: a global and a local contribution. 
A pristine halo is considered for Pop III star formation or treated as a DCBH candidate 
depending on the halo's virial temperature and the $J_{\rm LW}$ that it is exposed to.
\item The number density of DCBH sites can be up to $0.1\ \rm Mpc^{-3}$ at $z\!=\!6$, 
much higher than previously anticipated \citep{Dijkstra:2008p45}.
\label{DC}
\end{enumerate}

In the present study we refine the model of DCBH formation and also 
follow the subsequent growth of these seeds as identified in A12. We discuss new additions to the A12 model 
in the subsections below.

\subsection{DCBH forming haloes}

A DCBH forms in our model if a pristine massive halo is exposed to 
$J\geq J_{\rm crit}$ \footnote{Note that $J_{\rm crit}$ is the critical level of 
LW radiation required by a pristine atomic cooling halo to undergo direct collapse. 
The critical level of extragalactic LW radiation required by a pristine atomic cooling halo from Pop III stars, $\sim 1000$ and from Pop II stars $\sim 30-100$ 
\citep{WolcottGreen:2011p121}} \citep{WolcottGreen:2011p121} and satisfies 
both the spin and size criterion required for the disc to withstand fragmentation 
\citep[LN06 and LN07 hereafter]{Lodato:2006p375,Lodato:2007p869}. Note that the $J_{\rm crit}$ is always 
produced by stellar sources external to the pristine DCBH candidate 
halo as internal sources would pollute the gas inside the halo. 
We assume the collapsing gas in pristine haloes will settle into a disc 
whose stability against fragmentation determines whether it will be able 
to collapse to a BH or will fragment into star-forming clumps 
(LN06, LN07). Assuming that initially 
the baryons have the same specific angular momentum as the halo, 
the halo must have a spin, $\lambda$, lower than a characteristic value, $\lambda_{\rm max}$, 
for a given Toomre stability parameter $Q_c$, for which the pristine gaseous disc 
is exactly marginally stable and above which no accretion can take 
place onto the central region (LN06). The critical spin is given as
\begin{equation}
\lambda_{\rm max}= \frac{m_{\rm d}^2 Q_{\rm c}}{8j_{\rm d}} \sqrt{\frac{T_{\rm gas}}{T_{\rm vir}}} \ ,
\end{equation}
where $m_{\rm d}$ is the disc mass expressed as a fixed fraction (0.05) of 
the total baryonic mass in the halo \citep{Mo:1998p837}, $j_{\rm d}$ is the 
specific angular momentum of the disc that is also a fixed fraction (0.05) 
of the halo's overall angular momentum (LN06), $T_{\rm vir}$ is the virial 
temperature of the halo and $T_{\rm gas}$ is the temperature of the gas 
in the disc which depends on whether atomic or molecular hydrogen is the 
dominant cooling species. In our case, since the halo is exposed to $J\ge J_{\rm crit}$, 
the dominant coolant is atomic hydrogen and $T_{\rm gas}$ is set to 8000 K.
The second condition comes from the limit that the disc must be cooler than a characteristic temperature above which the gravitational torques 
required to redistribute the angular momentum become too large and can disrupt the disc. $T_{\rm max}$ is used as a proxy for size of the disc and is defined as 
\begin{equation}
T_{\rm max}=T_{\rm gas}\left(\frac{4\alpha_{\rm c}}{m_{\rm d}} \frac{1}{1 + M_{\rm BH}/m_{\rm d}M}\right)^{2/3}\ ,
\end{equation}
where $\alpha_{\rm c}$ is a dimensionless parameter (0.06) relating the critical 
viscosity to the gravitational torques in a halo with DM mass, $M$. This provides a 
mass estimate for the assembling DCBH (LN07)
\begin{equation}
M_{\rm BH} = m_{\rm d}M \left( 1 - \sqrt{ \frac{8 \lambda j_{\rm d}} {m_{\rm d}^2 Q_{\rm c} } \left(\frac{T_{\rm gas}}{T_{\rm vir}}\right)^{1/2}}\right)\ ,
\end{equation}
for $\lambda<\lambda_{\rm max}$ and $T_{\rm vir }< T_{\rm max}$.

We find that the inclusion of these two criteria for efficient angular momentum transport and 
accretion within the disc, in addition to our existing framework for the treatment of 
LW radiation feedback, fundamentally alters the progression of structure formation in these haloes, 
and impacts the observable characteristics of stars and the BHs within them. 

We plot the $T_{\rm vir}-\lambda$ distribution of pristine atomic cooling haloes 
that are exposed to $J_{\rm LW} \ge J_{\rm crit}$, for different values of $Q_c$ in Fig.~1. 
The haloes with spin in the range $\lambda < \lambda_{\rm max}$ are marked in red, 
with the almost-vertical solid curve representing $\lambda_{\rm max}$, whereas the 
ones with $\lambda > \lambda_{\rm max}$ are plotted in black.The size constraint in order for the disc to withstand fragmentation is denoted 
by the dashed curve. These limits together constrain 
DCBH formation to a small allowed domain in the  $T_{\rm vir}-\lambda$ plane marked by the yellow region.

Note that LN06, LN07 require the gas disc to be marginally stable i.e. $Q_c\sim O(1)$. Given that the actual high--redshift disc parameters are uncertain, we choose values of $Q_c$ close to unity and use $Q_c=3$ in our fiducial model, which sets an upper limit on the number of DCBHs with reasonable disc parameters and for which the disc sizes are not too large. This yields DCBHs (blue points) with a co-moving number density of 
$0.03\ \rm {Mpc}^{-3}$ in our fiducial case with $Q_c=3$, and $f_{\rm lim}=0.75$ (see the following section for $f_{\rm lim})$. 

\subsection{Star Formation}

In our model, Pop III stars form in pristine haloes subject to the following physical prescriptions/effects, discussed in A12 in more detail.
\begin{itemize}
\item Pop III star formation is prohibited due to LW feedback in pristine haloes with $2000 \leq \rm T_{\rm vir}< 10^4 \ \rm K$ even when J$_{\rm LW} < \rm J_{\rm crit}$ \citep{Machacek:2001p150,OShea:2008p41}. A pristine mini-halo that is subject to even a small value of LW radiation needs to be above a characteristic mass to host  a Pop III star due to the partial dissociation (and hence inefficient cooling) of H$_2$ molecules.
\item Pop III stars form following a top-heavy Salpeter IMF with mass limits dependent on halo's virial temperature, i.e. a single star with mass cut-offs at $[100,500] \ \msun$ in haloes with $2000 \leq \rm T_{\rm vir}< 10^4 \ \rm K$ and 10 stars with mass cut-offs at $[10,100] \ \msun$ in haloes with $T_{\rm vir}\geq 10^4 \ \rm K$. 
\end{itemize}

We consider a halo polluted if it has hosted a star or merged with a halo hosting a star.We set a mass threshold of $M>10^8\ \rm M_{\sun}$ for polluted haloes to form Pop II stars \citep{Kitayama:2004p669,Whalen:2008p785,Muratov:2012p2436}, following the reasoning that a halo needs to be massive enough to allow for the fall back or the retention of metals ejected from a previous Pop III star formation episode. In these polluted haloes, baryons are allowed to co-exist in the form of \textit{hot} non-star-forming gas, \textit{cold} star-forming gas, stars, or those locked into a DCBH that might have formed in or ended up in the halo through a merger.

We assume in our model that a DM halo is initially comprised of hot gas, $M_{\rm hot} = f_{\rm b} M_{\rm DM}$, where $f_{\rm b}$ is the universal baryon fraction and $M_{\rm DM}$ is the halo's current DM mass \footnote{Using a lower baryon fraction linearly affects the BH mass and evolution discussed in this work.}. We add non-star-forming gas to the halo by calculating the accretion rate, $\dot M_{\rm acc}$, defined as 
\begin{equation}
\dot M_{\rm acc} \equiv  \frac {f_{\rm b} \Delta M_{\rm DM} - M_{\rm *,p} - M_{\rm out,p} - M_{BH}}{\Delta t} \ ,
\label{eq.sam1}
\end{equation}
In this model $\Delta M_{\rm DM}$ is amount by which the DM halo grows between two snapshots separated by $\Delta t$ years. $M_{\rm *,p}$ and $M_{\rm out,p}$ represents the total stellar mass and net mass lost (from both cold and hot gas reservoir) in previous SN outflows at the beginning of the time step, respectively. $M_{\rm BH}$ is the total mass of the DCBH in the halo. 

The hot gas, $M_{\rm hot}$, converts into cold gas, $M_{\rm cold}$, by collapsing over the dynamical time\footnote{$t_{\rm dyn}$ is defined as the ratio of the halo's virial radius to the circular velocity defined for the infall mass and infall redshift} of the halo, $t_{\rm dyn}$. Pop II star formation can then occur via a Kennicutt-type relation \cite{Kennicutt:1998p143} 
\begin{equation}
\dot{M}_{*,\rm II}=\frac{\alpha}{0.1 t_{\rm dyn}}M_{\rm cold} \ ,
\label{eq.sam2}
\end{equation}
where $\alpha$ is the star formation efficiency set to 0.005 (SFE) and the factor $0.1t_{\rm dyn}$ is motivated by the angular momentum conservation condition for the central galaxy in a DM halo \citep[ and see A12 for a descriptions of the parameters used]{Kauffmann:1999p839, Mo:1998p837}. 

In Pop II star forming haloes, the outflow rate due to SN feedback is computed via the relation: $\dot{M}_{\rm out} =\gamma \  \dot{M}_{*,\rm II}$, where $\gamma = \left(\frac{V_{\rm c}}{V_{\rm out}}\right)^{-\beta}$ \cite{Cole:2000p863}.

We set $V_{\rm out}=110 \ \rm km\, s^{-1}$ and $\beta = -1.74$ resulting in typical values of $\gamma \approx 20$, following the results of the high resolution hydrodynamical simulations of the high redshift Universe (Dalla Vecchia and Khochfar 2013, in prep).
 
We track the evolution of baryons with the following set of coupled differential equations for the individual baryonic components:
\begin{equation}
\dot{M}_{\rm cold} = \frac{M_{\rm hot}}{t_{\rm dyn}} - \dot{M}_{*,\rm II} - \dot{M}_{\rm out} - \dot{M}_{\rm BH,cold}\ ,
\label{eq.sam3}
\end{equation}
\begin{equation}
\dot{M}_{\rm hot} =  \dot M_{\rm acc} -\frac{M_{\rm hot}}{t_{\rm dyn}} - \dot{M}_{\rm BH,hot} \ .
\label{eq.sam4}
\end{equation}
Equations \ref{eq.sam1}, \ref{eq.sam2},\ref{eq.sam3} and \ref{eq.sam4} are solved numerically over 100 smaller time steps between two snapshots.

\begin{figure}
\centering
\includegraphics[width=\columnwidth]{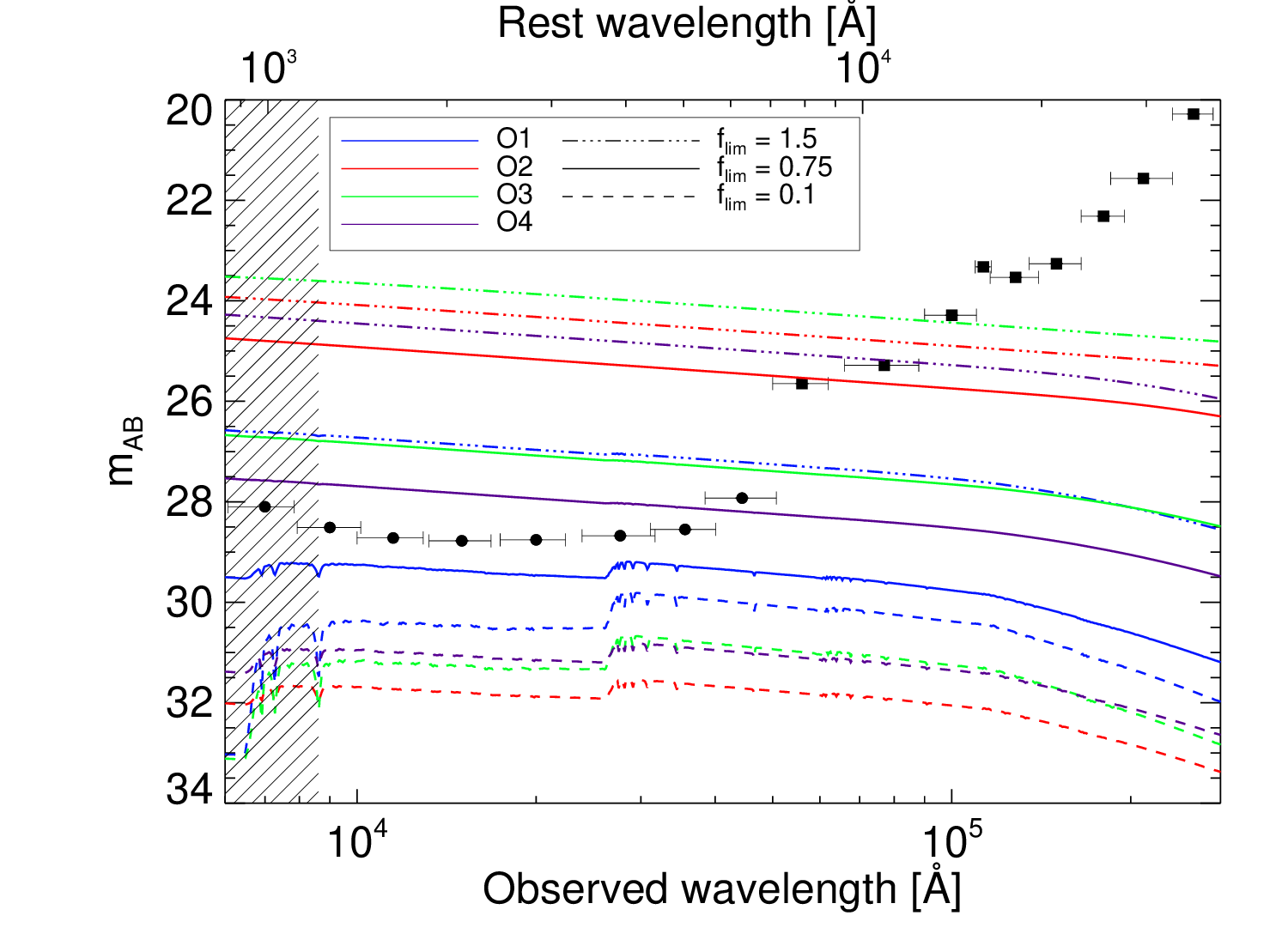}
\caption{OBG candidates and their observability with JWST at $z \sim 6$. 
We plot the observed flux density in AB magnitudes for all the OBGs, 
while varying the maximal allowed accretion rate, $f_{\rm lim}$. When the 
BH accretion dominates the spectrum there is virtually no Balmer break, and the UV slope is obviously 
fixed by the accreting black hole. The black points denote the flux limits and bandpass widths of NIRCam 
(circles) and MIRI (squares) wide filters, assuming a 10,000 second exposure with JWST and 
a S/N ratio of 10. The shaded area marks the wavelength region shortward Ly$\alpha$, 
where intergalactic neutral hydrogen is expected to completely absorb the OBG signal. }
\label{Figure3}
\end{figure}

\subsection{Growth of a DCBH}

Haloes hosting DCBHs are initially not massive enough and are not polluted enough to lead to Pop II star formation (Schneider et al. 2002). It is reasonable to assume that prior to the introduction of a stellar component, the gas reservoir is still massive enough to feed the central BH. At this stage the accreting DCBH might appear as a mini-quasar, but of essentially zero metallicity.
Following this epoch, however, the BH grows by accreting gas available in the halo, unchallenged by any further star formation until Pop II stars start forming in the halo, or until the halo merges with another halo hosting stars. The stellar component and the BH from this point on begin to grow in tandem, marking the onset of the OBG phase. How the two components evolve in detail is sensitive to the accretion rate and subsequent merging history - a parameter space that we have explored extensively.

Once a  DCBH forms in a halo, it is allowed to grow at a fixed fraction $f_{\rm lim}$ of the 
Eddington accretion rate, assuming that both the cold gas and hot gas can be accreted by the BH. The 
upper limit of the accretion rate is set by the parameter $f_{\rm lim}$ that we vary between individual runs. 
If the total gas available during our integration time steps for accretion is less than this 
fraction, the total mass available sets the accretion rate. We run our model for $f_{\rm lim}=[1.5,1.0,0.75,0.1]$ 
to explore the parameter space. From the model, at any given time step, $\Delta t$ (in Myr), the accretion efficiency computed from the gas reservoir is
\begin{equation}
f_{\rm model} = \ln \left( 1 + M_{\rm g}/M_{\rm BH}\right) \times \frac{ \epsilon}{(1-\epsilon)} \frac{\ 450 \  {\rm Myr}}{\Delta t} \ ,
\label{eq.f}
\end{equation}
where $M_{\rm g}$ is the total gas available in the halo at timestep $\Delta t$  and $M_{\rm BH}$ is the DCBH mass with the radiative efficiency, $\epsilon$, set to 10~$\%$. The accretion efficiency then used for the actual computation of the increase in the DCBH mass is
\begin{equation}
f_{\rm acc} = \min [f_{\rm model}, f_{\rm lim}]\ ,
\label{eq.lim}
\end{equation} 

Finally, we write
\begin{equation}
M_{\rm BH, final}=M_{\rm BH, ini} \exp\left(f_{\rm acc} \frac{1-\epsilon}{\epsilon}\frac{\Delta t}{450 \  {\rm Myr}}\right) ,
\end{equation}

Our fiducial case corresponds to a LW escape fraction of $1.0$, Pop II star formation
 efficiency of $0.005$, $Q_c=3$ and $f_{\rm lim}=0.75$. Note that the number of DC sites are directly dependant on $f_{\rm esc}$ and $\alpha$, where increasing the values of those parameters leads to a higher number of DC sites (A12). The number of DCBHs that form from those sites directly depends on $Q_{\rm c}$, where a higher value of $Q_{\rm c}$ leads to a higher number of DCBHs. The BH accretion parameters only affect the mass accreted by the DCBH as seen in Fig. \ref{Figure2} (see section \ref{sec:Results}). In haloes which host a DCBH but are not massive enough to form Pop II stars, 
the gas is assumed to be hot and diffuse (i.e. has not condensed over the dynamical time of the halo). 
In haloes which host a DCBH and a Pop II stellar component, both the hot and cold phases of gas are 
assumed to contribute to the accretion process. The total mass accreted by the DCBH 
is split into hot and cold components depending on the ratio of the hot and the cold gas reservoirs. 
We do not assume any feedback from the accreting DCBH affecting star formation in the galaxy. 
We do this to avoid inserting a correlation between the BH and stars by 
assuming such a feedback loop since the precise nature of accretion and 
feedback in galactic nuclei is largely unknown at $z>6$. 

To summarise, the total gas mass available for accretion, and hence the total 
mass accreted by the DCBH, depends on whether Pop II stars are forming in the halo or not. 
If there is no assembling Pop II stellar component, the DCBH is assumed to 
accrete from the hot gas reservoir, i.e the limiting accretion efficiency, $f_{\rm acc}^{\rm hot}$, is determined via eqs. \ref{eq.f} and \ref{eq.lim} using the hot gas ($M_{\rm hot}$ in eq. \ref{eq.f}) in the halo.
\begin{equation}
M_{\rm BH, final}=M_{\rm BH, ini} \exp\left(f_{\rm acc}^{\rm hot} \frac{1-\epsilon}{\epsilon}\frac{\Delta t}{450 \  {\rm Myr}}\right) ,
\end{equation}
If the halo hosts a Pop II stellar component, the limiting accretion efficiency, $f_{\rm acc}^{\rm hot + cold}$, is determined via eqs. \ref{eq.f} and \ref{eq.lim} using the hot and cold gas ($M_{\rm hot} + M_{\rm cold}$ in eq. \ref{eq.f}) in the halo.
\begin{equation}
M_{\rm BH, final}=M_{\rm BH, ini} \exp\left(f_{\rm acc}^{\rm hot+cold} \frac{1-\epsilon}{\epsilon}\frac{\Delta t}{450 \  {\rm Myr}}\right) ,
\end{equation}
The net BH mass accreted in a time step, $M_{\rm BH,acc} =M_{\rm BH, final}-M_{\rm BH, ini}$, can then be written as a sum of the hot ($M_{\rm BH, acc}^{\rm hot}$) and cold ($M_{\rm BH, acc}^{\rm cold}$) components
\begin{equation}
M_{\rm BH, acc} = M_{\rm BH, acc}^{\rm hot} + M_{\rm BH, acc}^{\rm cold}, \\
\end{equation}
The individual masses are computed as, 
\begin{eqnarray}
M_{\rm BH, acc}^{\rm cold}=RM_{\rm cold}, \\
M_{\rm BH, acc}^{\rm hot}=(1-R)M_{\rm hot}, \\
\end{eqnarray}
where $R = \frac{M_{\rm cold}}{M_{\rm hot}}$, if $M_{\rm cold}<M_{\rm hot}$, else $R = \frac{M_{\rm hot}}{M_{\rm cold}}$.
$M_{\rm BH, acc}^{\rm cold}$ and $M_{\rm BH, acc}^{\rm hot}$ are then used in eq. \ref{eq.sam3} and \ref{eq.sam4} to compute the updated hot and cold gas fractions.
\begin{figure}
\centering
\includegraphics[width= \columnwidth]{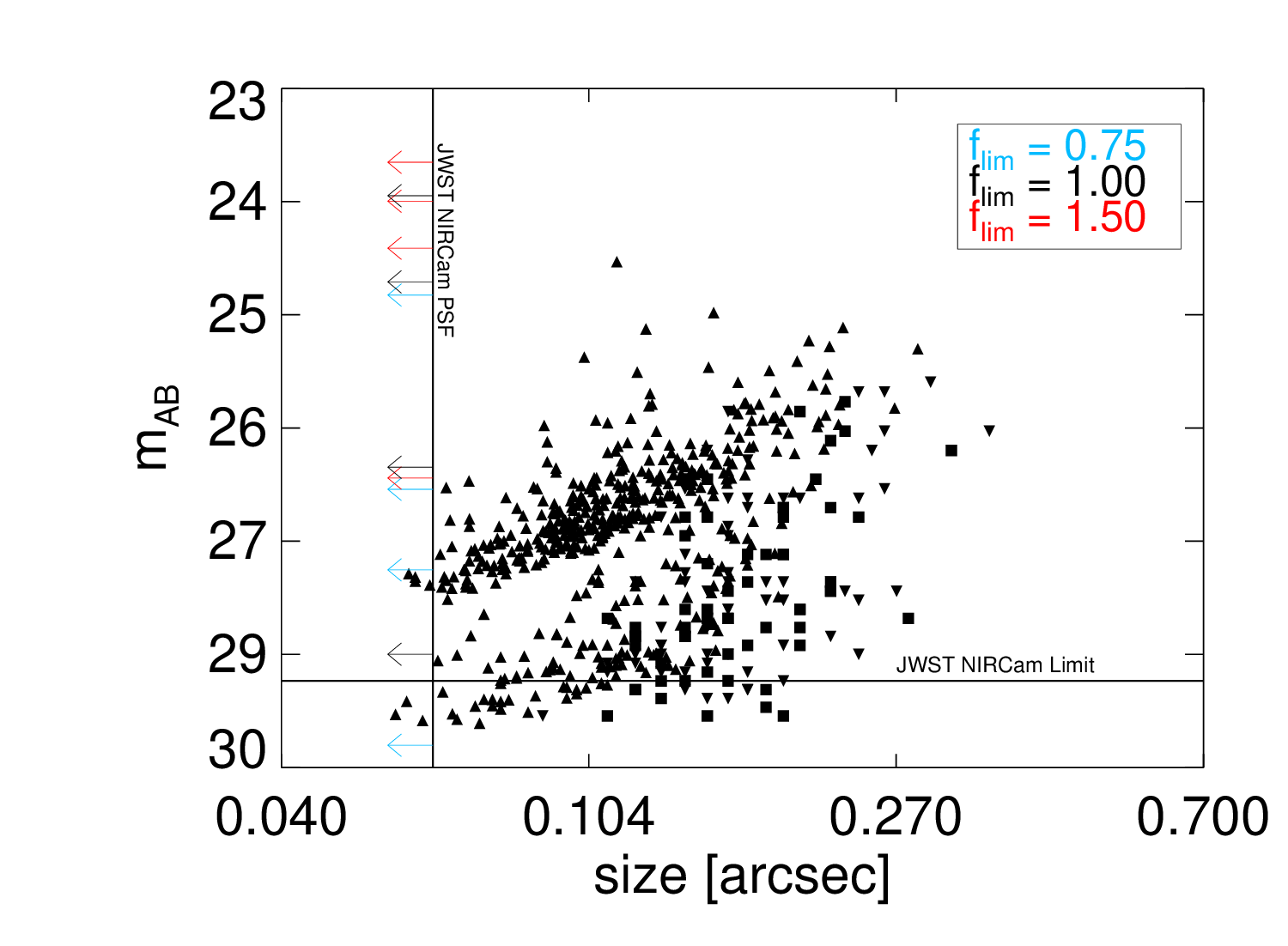}
\caption{Size versus magnitude relation. The HUDF galaxies, not corrected for point spread function (PSF), at $z=6,7,8$ are represented by upright triangles, 
downward triangles and squares respectively. Observational limits and the PSF of the NIRCam are plotted as the straight lines. The OBGs O1--O4 in our sample, 
denoted by the arrows pointing left, would be unresolved objects that could be brighter than the galaxies. Note that we have excluded the $f_{\rm lim}=0.1$ case from this plot as the m$_{\rm AB}$ for the OBGs is quite high. }
\label{Figure5}
\end{figure}
%
\begin{table}
\caption{Summary of cases considered in our work.}
\begin{tabular*}{0.45\textwidth}{@{\extracolsep{\fill}}lll}
\hline
Name &Symbol &Value \\
\\
\hline
Pop II star formation efficiency& $\alpha$ & 0.005 \\
LW escape fraction&$ f_{\rm esc}$& 1.0 \\
Radiative efficiency&$\epsilon$ & 0.1 \\
Limiting Eddington Accretion fraction& $f_{\rm lim}$ & 0.1-1.5 \\
Toomre Parameter& $ Q_c$ & 1.5-3\\ 
\hline
\end{tabular*}
\label{tablecases}
\end{table}
%

\section{Results}
\label{sec:Results}
For $Q_c = 3$, we find four OBGs, named O1--O4, in our simulation box. The stellar and black-hole growth tracks of these OBGs are shown in Fig.~2, 
colour coded as O1--blue, O2--red, O3--green, O4--purple, respectively. The fiducial case ($f_{\rm lim} =0.75$) is marked by the solid curves and the 
open triangles and circles denote the time--steps at $f_{\rm lim}=0.1,1.5$. The grey shaded region is where the BH would need to 
accrete at super--Eddington rates for the galaxy to appear as an OBG. Since these objects have $M_{\rm BH} > M_*$, the nuclear emission can 
dominate the starlight even when accreting at significantly sub-Eddington rates in the non--shaded region. 

Note that these OBGs preferably form in low mass atomic cooling haloes as seen in Fig.. 1. Almost all the DC candidate haloes meet the spin cut, but only the lower mass haloes, close to $10^7 \ \msun$, meet the size cut to allow for the formation of the DCBH. We also report that DCBH host haloes are in fact satellites of larger haloes hosting Pop II stars, in which the DCBH haloes eventually end up. This is, expected as the critical value of the LW radiation is generally produced by the larger star forming halos forming a close--pair with the DC candidate halo. The abundances of these objects at $z\sim 6$ is similar to the ones reported by \cite{Volonteri:2010p119}, for their 'low--threshold' case of DC seed formation.

\subsection{Observational predictions}
After identifying the sites of DC (A12), inclusion of physical processes that leads to their formation (LN06), enables the calculation of the observational signatures of OBGs. Haloes that harbour growing DCBH seeds with no (or little) associated stellar 
component merge into haloes that have formed the first and second generation of stars. We compute the observed spectral energy distribution (SED) of these copiously 
accreting DCBH seeds and the population of stars within the OBGs. To model the stellar component of the SED, we use the stellar masses and ages from our merger tree and 
derive the spectrum using \textsc{Starburst99} \citep{Leitherer:1999p112}. The accretion disc spectrum 
is modelled as a radiating blackbody with a temperature profile of the disc given by the alpha-disc model \citep{Pringle:1981p1672}. Since 
OBGs are expected to lie in haloes of low metallicity we do not include any dust absorption in our models.
Note that an OBG is characterised by possessing an actively accreting BH and an underlying stellar population that is Pop III or Pop II or both, however an OBG might appear as what has often been referred to as mini-quasars in the very early stages of its evolution when no stellar component is found in the DCBH host galaxy.

The UV-optical SED of an OBG is inevitably dominated by the accretion onto the central black hole. 
The predicted SEDs of the OBGs over the wavelength range observable with the NIRCAM (Near Infra-red Camera) and MIRI (Mid-Infra red Instrument) 
instruments aboard NASA's proposed {\it James Webb Space Telescope} (JWST) are shown in Fig. \ref{Figure3}. The stellar spectrum dominates the OBG spectrum 
only when the BH is limited to $< 0.1\ \rm f_{edd}$ at all times. However, such a low rate would be incompatible with the dramatic 
mass growth rates expected of the most massive, early black holes \citep[e.g.][]{Sijacki:2009p2208}.

We note that the magnitude of our brightest OBGs could be $\rm m_{AB} \approx 25$, comparable to the brightest putative Lyman-break galaxies uncovered at $z \simeq 7$ 
in ground-based surveys such as UltraVISTA \citep{Bowler:2012p2352}. However, it will be hard to distinguish OBGs from Lyman-break galaxies with ground-based 
imaging because, while the predicted UV continuum black-hole emission is expected 
to be relatively blue (with a UV slope $\beta \sim -2.3$, where $f_{\lambda} \propto \lambda^{\beta}$), it is not significantly bluer than that displayed by the 
general galaxy population at these early times \citep[e.g.][]{Dunlop:2012p2380}. OBGs are also of course expected to display a negligible Balmer break. However, 
while it is interesting that the stack of the brightest Lyman-break galaxies in UltraVISTA shows at most a very weak Balmer break, it will still require
extremely high-quality mid-infrared photometry to conclusively rule out the possibility that the UV-optical SED of a putative Lyman-break 
galaxy is incompatible with that produced by a very young stellar population.

Identification of OBGs amidst the observed high-redshift galaxy population will therefore require high-resolution imaging with HST and ultimately 
JWST. With $M_{\rm BH} > M_*$, an OBG accreting at a reasonable fraction of the Eddington limit should certainly appear unresolved and point-like
in high-resolution rest-frame UV (observed near-infrared) imaging. As illustrated in Fig. 4, it is already known that the vast majority of 
{\it faint} high-redshift 
galaxies uncovered via deep HST imaging are resolved \citep[see][]{Oesch:2010p2399,Ono:2012p2382}, but this does not rule out the existence of an OBG 
population with a surface density $< 1$ arcmin$^{-2}$, and HST follow-up of the brighter and rarer high-redshift ($z \simeq 7$) objects  uncovered by the 
near-infrared ground-based surveys is required to established whether or not they are dominated by central black-hole emission.

\section{Discussion and Conclusion}

In this study, we report the possible existence of OBGs, at $z>6$ in which the DCBH precedes the epoch of stellar assembly and outshines the stellar component in for a considerable fraction of the galaxy's lifetime. Our 3.41 Mpc $h^{-1}$ box produces about 4 of these OBGs. Although this is not a cosmological average owing to the small box size, the main aim of this study is to discuss the physical conditions that could lead to the existence of OBGs, which are effects that operate on a scale of less than a few tens pf physical kpc, mostly insensitive to our chosen box-size.

Besides the observational features discussed, like all active galactic nuclei, OBGs are expected to display broad-line emission from highly excited species in the vicinity of the black hole. However, as OBGs have very low metallicity, it is unclear whether lines such as NV and CIV 
are expected to be detectable even given high-quality near-infrared spectroscopy, and Lyman-$\alpha$ is often severely 
quenched by neutral Hydrogen as we enter the epoch of reionisation. 

Thus, the best observational route to establishing whether OBGs exist, and if so constraining their number density (and ultimately their 
evolving luminosity function), appears to be via deep imaging of putative high-redshift Lyman-break galaxies, 
with sufficient angular reolution to prove they are unresolved, coupled with sufficiently accurate photometry to 
prove any point-like objects cannot be dwarf star contaminants \citep[see, for example ][]{Dunlop:2012p2381}.

The discovery of an OBG could in principle settle the long standing debate on whether DCBHs can form and be the seeds of the first SMBHs \citep[A12,][]{Dijkstra:2008p45,Volonteri:2008p1043,Volonteri:2009p1399, Johnson:2012p2056, Bellovary:2011p1119}. Uncovering this population holds great promise for understanding the onset of black hole and host galaxy growth.

\section*{Acknowledgements}
The authors would like to thank Rychard Bouwens for providing the HUDF data used in Plot. \ref{Figure5}.  BA would like to thank Jarrett L. Johnson for his useful comments on the first draft. JSD acknowledges the support of the European Research Council via an Advanced Grant, and the support of the
Royal Society through a Wolfson Research Merit award.

\bibliographystyle{mn2e}


\label{lastpage}

\end{document}